# Gapped quantum spin liquid in a triangular-lattice Ising-type antiferromagnet PrMgAl$_{11}$O$_{19}$

Chengpeng Tu[1,2], Zhen Ma[3,1†], Hanru Wang[1], Yihan Jiao[1], Dongzhe Dai[1], and Shiyan Li[1,2,4,5*]

[1]*State Key Laboratory of Surface Physics, and Department of Physics, Fudan University, Shanghai 200438, China*

[2]*Shanghai Research Center for Quantum Sciences, Shanghai 201315, China*

[3]*Hubei Key Laboratory of Photoelectric Materials and Devices, School of Materials Science and Engineering, Hubei Normal University, Huangshi 435002, China*

[4]*Shanghai Branch, Hefei National Laboratory, Shanghai 201315, China*

[5]*Collaborative Innovation Center of Advanced Microstructures, Nanjing 210093, China*

Corresponding author. Email: zma@hbnu.edu.cn (Z.M.); shiyan_li@fudan.edu.cn (S.Y.L.)

## Abstract

**In the search of quantum spin liquid (QSLs), spin-1/2 triangular-lattice Heisenberg antiferromagnets (TLHAFs) have always been viewed as fertile soils. Despite the true magnetically-ordered ground state, anisotropy has been considered to play a significant role in stabilizing a QSL state. However, the nature and ground state of the most anisotropic case, the triangular-lattice Ising antiferromagnet (TLIAF), remains elusive and controversial. Here, we report specific heat and thermal conductivity measurements on a newly-discovered Ising-type QSL candidate PrMgAl$_{11}$O$_{19}$. At zero field, the magnetic specific heat shows a quadratic temperature dependence. On the contrary, no direct positive magnetic contribution to thermal conductivity was detected, ruling out the presence of mobile gapless fermionic excitations. Further analysis of phonon thermal conductivity reveals that the phonons are strongly scattered by thermally-activated magnetic excitations out of a gap, which exhibits a linear dependence with magnetic field. These results demonstrate that the spin-1/2 TLIAF PrMgAl$_{11}$O$_{19}$ has a gapped $Z_2$ QSL ground state.**

**Introduction**

Since the first proposal by Anderson in the 1970s[1], the pursuit of quantum spin liquids (QSLs) has been one of the major issues in current condensed-matter physics research[2-4], due to the exotic physical properties, potential connectivity with high-temperature superconductivity[5] and application for topological quantum computation[6,7]. In contrast to conventional ordered magnets, QSLs are highly entangled quantum states without spontaneous symmetry breaking and the formation of long-range magnetic order down to absolute zero temperature. Interestingly, fractionalized excitations like spinons can emerge from the QSL states and this characteristic was regarded as the core of the experimental detection and verification of QSL candidates[2-4].

Historically, spin-1/2 triangular-lattice Heisenberg antiferromagnet (TLHAF) was the first conceptual model in realizing QSL state due to the simplest geometrically frustrated spin configuration[1]. Although subsequent studies have identified the true ground state as a noncollinear 120° antiferromagnetic ordered state[8,9], the long-range order can be destructed by introducing additional perturbations such as next-nearest-neighbor interaction[10,11], ring interaction[12,13], anisotropic interaction[14-17], exchange randomness[18-20] or together. Experimentally, a series of QSL candidates have emerged based on two-dimensional (2D) triangular lattice, including *κ*-(BEDT-TTF)$_2$Cu$_2$(CN)$_3$ (ref. [21,22]), EtMe$_3$Sb[Pd(dmit)$_2$]$_2$ (ref. [23]), *κ*-H$_3$(Cat-EDT-TTF)$_2$ (ref. [24]), YbMgGaO$_4$ (ref. [25-27]), NaYbSe$_2$ (ref. [28,29]), TbInO$_3$ (ref. [30,31]) and YbZn$_2$GaO$_5$ (ref. [32]). Thereinto, exchange randomness is an inevitable factor in real materials and can lead to a mimicry of gapless QSL state: random-singlet state, the origin of which could be of variety, such as the slowing down of charge or proton degrees of freedom in these organic compounds, quenched disorder of random site occupancies of Mg$^{2+}$/Ga$^{3+}$ in YbMgGaO$_4$, and Zn$^{2+}$/Cu$^{2+}$ in kagome-lattice ZnCu$_3$(OH)$_6$Cl$_2$ (ref. [18-20]). Leaving the disorder effect behind, the exchange anisotropy in TLHAFs was argued to lay a foundation for enhancing the quantum fluctuations and thus stabilizing the QSL state[14-17]. For organic compounds, the anisotropy mainly stems from the spatially non-equivalent bond lengths (transfer integrals) of the isosceles triangular lattice[3,14,15]. While for inorganic especially rare-earth compounds, due to the strong correlation and much localized 4*f* electrons, the strong spin-orbit coupling contributes to highly anisotropic exchange interactions, and hence the generic spin Hamiltonians of TLHAFs are usually modified to *XXZ* model[25-32]. In fact, anisotropic exchange interactions also play a vital significance in QSL candidate with other lattice geometries like kagome-lattice $ZnCu_3(OH)_6Cl_2$ (ref. [33]), pyrochlore-lattice $Yb_2Ti_2O_7$ (ref. [34]) and Shastry-Sutherland-lattice $Pr_2Ga_2BeO_7$ (ref. [35]). From this perspective, one would naturally raise a question of whether the most anisotropic case, the Ising-type interaction with easy-axis along *c* direction, could also breed a QSL state.

Theoretically, a classical spin liquid ground state with finite zero-point entropy was firstly predicted in the classical triangular-lattice Ising antiferromagnet (TLIAF) by Wannier[36]. However, the macroscopic degeneracy will be lifted when introducing additional quantum fluctuations, which makes the true ground state of spin-1/2 TLIAFs vague. The debates remain up to now and vary from magnetically-ordered state[37,38] to possible QSL state[39]. In practice, the most common way to generate quantum fluctuations is applying additional traverse field to Ising spins, for instance, the transverse-field Ising model for 1D spin chain[40]. While in 2D TLIAFs, a more intrinsic mechanism is the crystal-electric-field (CEF) non-Kramer doublet ground state, in which the longitudinal spin component behaves as the magnetic dipole moment (Ising interactions) while the transverse spin component behaves as the magnetic quadrupole moment (quantum fluctuations)[41]. However, the experimental realization of spin-1/2 TLIAFs has been scarce so far, only in $NdTa_7O_{19}$ (ref. [42]), $TmMgGaO_4$ (ref. [43]) and $KTmSe_2$ (ref. [44]). Thereinto, the lack of $NdTa_7O_{19}$ single crystal blurs further studies[42], and $TmMgGaO_4$ shows a partial magnetic order below 0.7 K (ref. [43]).

Recently, a new class of 2D rare-earth triangular-lattice frustrated magnets $Pr(Mg/Zn)Al_{11}O_{19}$ was successfully synthesized and characterized[45-50]. While there are only polycrystals for $PrZnAl_{11}O_{19}$, the single crystals are available for $PrMgAl_{11}O_{19}$ (ref. [48-50]), beneficial to more detailed investigations. As shown in Fig. 1a,b, $PrMgAl_{11}O_{19}$ crystallizes in the space group *P6$_3$/mmc*, in the structure of which magnetic $PrO_{12}$ tetrakaidecahedrons construct ideal triangular layers with an *ABAB* stacking arrangement. Interlinked between the triangular layers are two sheets of nonmagnetic layers consist of $AlO_6$ octahedra and $Mg/AlO_4$ tetrahedra. Due to the large ionic radius difference in $Pr^{3+}$ and $Al^{3+}/Mg^{2+}$ ions, the random site occupancy between magnetic and nonmagnetic layers like in $ZnCu_3(OH)_6Cl_2$ is prohibited[45,48-50]. Moreover, compared with the well-studied $YbMgGaO_4$, the distance between the magnetic layers is relatively larger and the disordered $Mg/AlO_4$ layer is not directly adjacent to the magnetic layers, which means the structural disorder has less influence on the coordination environment of $Pr^{3+}$ ions and spin exchange interaction.

No long-range order was observed in $Pr(Mg/Zn)Al_{11}O_{19}$ by specific heat and ac susceptibility down to 50 mK, despite the negative Curie-Weiss temperatures $\Theta_c$ = -8.1 K/-8.9 K, respectively[47-50]. The frequency-independent ac susceptibility of Pr(Mg/Zn)Al$_{11}$O$_{19}$ favors persistent spin dynamics and excludes any spin freezing or spin-glass behavior like in Yb(Mg/Zn)GaO$_4$ (ref. [51]). Besides, inelastic neutron scattering (INS) study detected a broad continuum and the magnetic specific heat exhibited a nearly $T^2$ dependence, suggesting Pr(Mg/Zn)Al$_{11}$O$_{19}$ to be gapless QSL candidates[47,48]. More interestingly, as plotted in Fig. 1d, the field dependence of dc magnetization result shows a negligible magnitude along $a$ axis, in sharp contrast with that along $c$ axis, indicating the highly easy-axis anisotropy of PrMgAl$_{11}$O$_{19}$, consistent with previous reports[48-50]. The CEF analyses and electron spin resonance (ESR) results further identified the ground state of non-Kramer Pr$^{3+}$ ions as a quasidoublet with effective $S$ = 1/2 (ref. [48,49]). These features make PrMgAl$_{11}$O$_{19}$ an ideal platform to study possible QSL state in spin-1/2 TLIAF.

In this paper, we report the specific heat and thermal conductivity measurements on high-quality PrMgAl$_{11}$O$_{19}$ single crystals. At zero field, a large magnetic contribution with a quadratic temperature dependence is observed in the specific heat. However, no direct positive contribution from magnetic excitations is detected in the thermal conductivity, indicating the absence of mobile gapless fermionic excitations. On the other hand, a clear bifurcation deviated from pure phonon thermal conductivity is observed at all fields, indicating strong scattering by thermally-activated magnetic excitations out of a gap. By analyzing the spin-phonon scattering rate, we estimate this tiny gap $\Delta$ ~ 0.153 K at zero magnetic field, which increases linearly with applied field. The spin-phonon scattering rate satisfies a unified form $1/\tau_{sp}$ ~ $T\exp(-\Delta/T)$, demonstrating a gapped $Z_2$ QSL ground state in PrMgAl$_{11}$O$_{19}$.

## Results

Figure 2a shows the specific heat of PrMgAl$_{11}$O$_{19}$ single crystal at various magnetic fields up to 11 T in the temperature range of 0.3 to 3 K. After subtracting the zero-field data of nonmagnetic counterpart LaMgAl$_{11}$O$_{19}$ polycrystalline sample as phonon contribution, the magnetic specific heat $C_m$ of PrMgAl$_{11}$O$_{19}$ is extracted and plotted in Fig. 2b in a log-log scale. At zero field, the low-temperature $C_m$ below 1.5 K shows a nearly power-law behavior, and can be well fitted to $C_m = cT^\beta$ with $\beta$ = 1.996 ± 0.014. The quadratic temperature dependence is consistent with previous results of Pr(Mg/Zn)Al$_{11}$O$_{19}$ (ref. [47,48]), reminiscent of a gapless $U(1)$ Dirac QSL state in which $C_m$ ~ $T^2$ is theoretically predicted due to the linear dispersion with nodal structure of spinon spectrum[52]. In fact, this feature has also been observed in other QSL candidates, for instance, YbZn$_2$GaO$_5$ (ref. [32]), YCu$_3$(OH)$_{6.5}$Br$_{2.5}$ (ref. [53,54]) and Pr$_2$Ga$_2$BeO$_7$ (ref. [35]).

The picture is rather complicated under magnetic fields. As the field increases, the magnetic specific heat is rapidly suppressed and the temperature dependence transforms from a power-law behavior to an exponential one, indicative of the opening of spin gap. However, beyond 1.5 T, it is worth noting that the $C_m$ seems to switch back to a nearly $T^2$ power-law behavior. The switching temperature gradually increases with the magnetic field and exceeds 3 K at 11 T, which finally becomes a $T^2$ behavior in the whole temperature range of our measurement, as shown in Fig. 2b. Such an unusual feature at high fields was also observed in PrZnAl$_{11}$O$_{19}$, which was ascribe to the inhomogeneous effective magnetic field for polycrystals due to the random crystallographic axis orientations of grains[47]. Our single crystal study firmly excludes this factor. Experimentally, the $T^2$ magnetic specific heat in disordered systems was also explained by a high degree of 2D magnetic order[55] or gapless linearly dispersing Halperin-Saslow modes[56]. However, our subsequent thermal conductivity measurements above 3 T do not observe any significant contribution from magnetic excitations. Since another ultralow-temperature specific heat study has observed evident Schottky anomalies[50], a trivial explanation is that the $T^2$ behavior at high fields may just originate from a crossover region from gapped exponential behavior to a high-temperature tail of Schottky contribution $C_{Schottky} \sim T^{-2}$. We leave it as an open question here and call for further investigation.

To further investigate the nature of low-energy magnetic excitations of PrMgAl$_{11}$O$_{19}$, we performed ultralow-temperature thermal conductivity measurement, a clean and powerful technique in studying the low-energy magnetic excitations in QSL candidates[27,29,57-59]. Comparing with specific heat measurement, thermal conductivity is insensitive to Schottky anomalies and CEF excitations, which is more beneficial to identify the ground nature. On the one hand, it can directly detect the positive contribution from mobile gapless fractionalized excitations, for instance, in BaCo$_2$(AsO$_4$)$_2$ (ref. [57]) and PbCuTe$_2$O$_6$ (ref. [58]). On the other hand, in most instances, the magnetic excitations in quantum magnets make a negative contribution by scattering phonons rather than conducting heat[27,29,59,60]. By analyzing the scattering rate, one could obtain information on the excitation spectrum[59]. For an insulating paramagnet especially QSL candidate, the thermal conductivity at low temperatures can usually be decomposed as $\kappa = aT + bT^\alpha$, where the two terms $aT$ and $bT^\alpha$ represent contributions from gapless mobile fermionic excitations (if exist) and phonons, respectively[61,62]. For phonons, the power $\alpha$ is typically between 2 and 3 due to the specular reflections at the sample surfaces[61,62].

The in-plane ultralow-temperature thermal conductivity of PrMgAl$_{11}$O$_{19}$ single crystal in zero and various magnetic fields up to 5 T is displayed in Fig. 3a. At high fields like 3 T and 5 T, the thermal conductivity data get saturated and independent of magnetic field below 0.8 K, indicating pure phonon contribution. In fact, the fitting to $\kappa/T = a + bT^{\alpha-1}$ for 5 T data below 0.3 K yields $\kappa_0/T \equiv a = -0.010 \pm 0.003$ mW K$^{-2}$ cm$^{-1}$ with $\alpha = 2.48 \pm 0.03$, as plotted in Fig. 3c. Considering our typical experimental error bar $\pm 0.005$ mW K$^{-2}$ cm$^{-1}$, the negligible residual linear term $\kappa_0/T$ at 5 T is virtually zero, reinforcing that the thermal conductivity is solely contributed by phonons without being scattered by magnetic system. With decreasing field, although the thermal conductivity data still overlap with the high-field curves below certain temperatures $T_s$, they are suppressed more and more strongly above $T_s$. To see more details, $\Delta(\kappa/T) \equiv (\kappa(H) - \kappa(5\ T))/T$ is extracted from raw data and the overlapped regions are clearly shown in Fig. 3b. The arrow in Fig. 3b denotes the corresponding $T_s$ at each field, which is defined as the temperature when $\Delta(\kappa/T)$ starts to exceed -0.01 mW K$^{-2}$ cm$^{-1}$, twice our experimental error bar. Such a suppression temperature $T_s$ has also been observed in other quantum magnets like Pr$_2$Ir$_2$O$_7$ (ref. [63]), CoNb$_2$O$_6$ and NiNb$_2$O$_6$ (ref. [40]), indicating additional scattering of phonons by magnetic excitations out of a small gap. At finite fields, the existence of thermally-excited low-energy magnetic excitations out of a gap is consistent with specific heat results. These magnetic excitations contribute to specific heat, and scatter phonons thus reduce the phonon thermal conductivity.

At zero field, nevertheless, it is hard to distinguish whether an overlapped region with pure phonon thermal conductivity, namely, a finite $T_s$ exists, as shown in Fig. 3c. According to the magnetic specific heat results of PrMgAl$_{11}$O$_{19}$, for a $U(1)$ Dirac QSL scenario, one would expect a significant contribution to residual linear term by gapless spinons since the unavoidable impurity scattering in real materials will broaden the nodes and produce a finite density of states (DOS) at low energy, analogous to the case of $d$-wave superconductors[64]. In fact, a recent heat transport study on Pr$_2$Ga$_2$BeO$_7$ has observed both $C_m \sim T^2$ and finite $\kappa_0/T$ at 0 T (ref. [35]). Nevertheless, as sketched in Fig. 3c, the fitting to $\kappa/T = a + bT^{\alpha-1}$ for 0 T below 0.3 K of PrMgAl$_{11}$O$_{19}$ yields $\kappa_0/T \equiv a = -0.061 \pm 0.005$ mW K$^{-2}$ cm$^{-1}$ with $\alpha = 1.68 \pm 0.02$. A negative $\kappa_0/T$ has no physical meaning, and the power is abnormally lower than 2, similar to YbMgGaO$_4$ (ref. [27]) and NaYbSe$_2$ (ref. [29]), which can be ascribed to strong spin-phonon scattering (scattering by magnetic system). This result indicates the absence of mobile gapless fermionic excitations in $PrMgAl_{11}O_{19}$.

To settle the discrepancy between specific heat and thermal conductivity measurements, a promising scenario is that $PrMgAl_{11}O_{19}$ has a gapped ground state. In fact, as plotted in Fig. 4c, the suppression temperature $T_s$ has a nearly linear dependence versus magnetic field. Most importantly, the linear fitting yields finite intercept, namely, $T_s = 0.055 \pm 0.005$ K at 0 T, which is below our measurement limit; hence, it is naturally to speculate that the thermal conductivity for 0 T and 5 T would overlap at lower temperature. This indicates the existence of a tiny spin gap at 0 T and explains the strongly scattering of phonons by thermally-activated magnetic excitations. Indeed, we notice that two recent specific heat measurements down to 50 mK at 0 T both observed a transformation from a $C_m \sim T^2$ behavior to an exponential-like one below about 200 mK (ref. [49,50]), implying an opening of spin gap at low temperature.

Having established that the magnetic excitations have no direct positive contribution to heat conduction, the thermal conductivity of $PrMgAl_{11}O_{19}$ is all contributed by phonons scattered by magnetic excitations. To gain more information about the nature of magnetic excitations in $PrMgAl_{11}O_{19}$, we now seek to quantify the spin-phonon scattering rate, following the method in ref. [59,65,66]. According to Matthiessen's rule, the total scattering rate of phonons is the sum of all independent origins $\tau^{-1}(T,H) = \tau_0^{-1} + \tau_p^{-1}(T) + \tau_{sp}^{-1}(T,H)$, where $\tau_0$, $\tau_p$ and $\tau_{sp}$ represent the boundary scattering, non-magnetic intrinsic phonon scattering (for instance, specular reflection) and additional spin-phonon scattering relaxation time, respectively. Since $\kappa$ is saturated beyond 3 T, the thermal conductivity data at highest field 5 T can be regarded as pure phonon contribution $\kappa_{ph}$. According to the kinetic formula, thermal conductivity at low temperature can be described as $\kappa_{ph}(T,H) = \frac{1}{3}C_{ph}v_{ph}l_{ph}(T,H) \propto T^3\tau(T,H)$. Thus, combining the above equations, the spin-phonon scattering rate $1/\tau_{sp}$ is proportional to $T^3(1/\kappa_{ph}(H)-1/\kappa_{ph})$. In Fig. 4a, we plot the $1/\tau_{sp}$ versus temperature at low fields above their corresponding suppression temperatures $T_s$.

To estimate the magnitude of the spin gap, we fitted $1/\tau_{sp}$ to an exponential function with a pre-factor function $f(T) = T^n$ like what nuclear magnetic resonance (NMR) study always did, $1/\tau_{sp} \sim T^n \exp(-\Delta/T)$, as shown in Fig. 4a. The obtained fitting parameters of power $n$ and spin gap $\Delta$ are plotted in Fig. 4b,c, respectively. Strikingly, the power $n$ is close to unity at all fields. This universal power behavior indicates that the magnetic excitations at zero field and finite fields are of the same origin and the spin-phonon scattering rate $\tau_{sp}^{-1}$ is proportional to $T$ if the excitation spectrum are gapless. In fact, the $T$-linear spin-phonon scattering rate has been theoretically predicted in 3D QSLs with $U(1)$ gauge[67], and a $\tau_{sp}^{-1} \sim T^{1.3}$ behavior with power close to unity has been observed in 2D QSL candidate $\kappa$-(BEDT-TTF)$_2$Cu$_2$(CN)$_3$ by magnetocaloric-effect measurements[68]. For comparison, the electron-phonon scattering obeys $\tau_{ep}^{-1} \sim T$ (ref. [69]) and the spin-phonon scattering of gapless AFM magnons satisfies $\tau_{sp}^{-1} \sim T^5$ (ref. [70]) at low temperature. The huge difference may imply that the magnetic excitations of PrMgAl$_{11}$O$_{19}$ are subject to Fermi-Dirac statistics, likely factionalized spinons. We notice that a similar behavior of Knight shift $^{17}K \sim T\exp(-\Delta/T)$ has been observed by NMR study of ZnCu$_3$(OH)$_6$Cl$_2$, in which the $T$-linear dependence was explained by the Pauli susceptibility of gapped fermionic spinons[71]. In addition, the obtained spin gap $\Delta \sim 0.153 \pm 0.010$ K at 0 T, and it shows a nice linear dependence versus magnetic field, the same as $T_s$. Note that the size of $\Delta$ is about 4 to 5 times of $k_BT_s$, which is a reasonable value since the thermally-activated excitations are usually capable of scattering phonons when $k_BT$ is about ~ 20% of the gap. Furthermore, the linearly-increasing gap with field may link with a magnetic Zeeman effect which yields $\Delta(H) = \Delta(0) + g_c\mu_BSH$, where $g_c$ is the Laudé g factor along $c$ axis and $\mu_B$ is the Bohr magneton. It is expected that $S = 1/2$ for spinons, whereas $S = 1$ for visons or spin triplet excitations[71,72]. According to previous ESR result that $g_c = 5.1$ for Pr$^{3+}$ (ref. [49]), our linear fitting of $\Delta(H)$ gives $S = 0.44$, which is near to 1/2, indicating spinon excitations.

## Discussion

Based on the analyses of our thermal conductivity results, the true ground state of PrMgAl$_{11}$O$_{19}$ is more inclined to be a gapped $Z_2$ QSL, in which the magnetic excitation spectrum has the form of $E \sim |k| + \Delta(H)$. The linear dispersion relation and the presence of a tiny gap can account for the $T^2$ magnetic specific heat while more exponential-like behavior below 200 mK (ref. [48-50]). Besides, such a tiny gap can be masked by the high intensity of incoherent scattering below 0.2 meV and instrumental resolution in the INS experiments, in which a gapless board continuum was observed[48]. To sum up, the gapped $Z_2$ QSL scenario matches and unifies all the experimental results of specific heat, thermal conductivity and INS well.

Indeed, the gapped $Z_2$ QSL can live in a neighborhood of the $U(1)$ Dirac QSL[73,74]. Hence, a $U(1)$

Dirac QSL can be regarded as the parent of gapped $Z_2$ QSL where the pairing amplitudes are turned off[73,74]. This gapless state is sensitive to spinon pairing instability, which can break the $U(1)$ gauge symmetry down to $Z_2$ and open a gap[74,75]. A common way to introduce instability is the application of external field. Theoretically, an infinitesimal field can destabilize the $U(1)$ Dirac QSL state and cause a spontaneous spin ordering with fully gapped spinons on the kagome lattice[76]. Besides, for Kitaev honeycomb lattice, the gapless $Z_2$ $B$ phase with Dirac spinons will also immediately acquire a gap when applying external field[6]. Indeed, such field-induced instabilities have been observed in gapless QSL candidates like EtMe$_3$Sb[Pd(dmit)$_2$]$_2$ (ref. [77]) and ZnCu$_3$(OH)$_6$SO$_4$ (ref. [78]). While for PrMgAl$_{11}$O$_{19}$, since the ground state at zero field is still gapped, the formation of gap has other mechanism, which may originate from the Ising anisotropy. Theoretically, a gapped $Z_2$ QSL has been proposed by density-matrix renormalization group (DMRG) studies as the ground state of spin-1/2 TLHAF with next-nearest-neighbor interaction[10,79,80] or spatial anisotropy[14]. As mentioned above, the exchange anisotropy due to strong spin-orbit coupling can also play the same role in stabilizing a QSL state. From this perspective, one would also expect a gapped $Z_2$ QSL ground state of the spin-1/2 TLIAF, warranting future study.

Experimentally, some other QSL candidates have also been proposed to have a gapped ground state, including $\kappa$-(BEDT-TTF)$_2$Cu$_2$(CN)$_3$ (ref. [22]), NaYbSe$_2$ (ref. [81,82]), ZnCu$_3$(OH)$_6$Cl$_2$ (ref. [71]) and YCu$_3$(OH)$_{6.5}$Br$_{2.5}$ (ref. [59]), albeit with debates[21,28,29,83-86]. For a gapped $Z_2$ QSL, two crucial issues are of vital significance. One is the classification of detected elementary quasiparticle excitations, which contain fermionic spinons, bosonic spinons and bosonic visons[74]. For $\kappa$-(BEDT-TTF)$_2$Cu$_2$(CN)$_3$, the observation of thermally-activated spin excitations in zero and finite fields up to 10 T by thermal conductivity[22] was ascribed to bosons[87]. (Note that the most recent understanding of $\kappa$-(BEDT-TTF)$_2$Cu$_2$(CN)$_3$ favors a gapped valence-bond-solid ground state and the analysis of thermal conductivity has been reconsidered[83].) And for ZnCu$_3$(OH)$_6$Cl$_2$, the NMR and INS measurements identified gapped fermionic spinons and visons, respectively[71,88]. While for PrMgAl$_{11}$O$_{19}$ here, since the prefactor function of spin-phonon scattering rate satisfies $1/\tau_{sp} \sim T$, analogy to electron-phonon scattering, the dominant scatters may be fermionic spinons with a linear dispersion $E \sim |k| + \Delta(H)$. The remaining unsolved issue is the field-evolution of $\Delta(H)$, which is diverse in different materials, including the field-independence of $\Delta(H)$ in $\kappa$-(BEDT-TTF)$_2$Cu$_2$(CN)$_3$ (ref. [22]), negative linear dependence of $\Delta(H)$ in ZnCu$_3$(OH)$_6$Cl$_2$ (ref. [71]) and YCu$_3$(OH)$_{6.5}$Br$_{2.5}$ (ref. [59]), and positive linear dependence of $\Delta(H)$ in PrMgAl$_{11}$O$_{19}$ here. The diversity may stem from different geometrical configuration, anisotropy interaction, and even field-induced magnetic transition. The unique field dependence of PrMgAl$_{11}$O$_{19}$ may imply the significance of Ising anisotropy in determining the structure of excitation spectrum, which will undoubtfully promote the understanding of gapped $Z_2$ spin liquid as well as spin-1/2 TLIAF.

Finally, despite the above discussions, there is still little room for the gapless scenarios. An indistinguishable case is that PrMgAl$_{11}$O$_{19}$ does have a $U(1)$ Dirac QSL state while the gapless spinons do not conduct heat, which may due to the localization effect by disorders. In fact, for PrMgAl$_{11}$O$_{19}$, a recent work has unveiled the presence of about 7% quenched disorder at the Pr site by single crystal structure refinement[49]. Beyond that, as mentioned above, disorders can also act as a role in inducing a random-singlet state[18-20]. This randomness-induced QSL-like state will mimic the gapless features such as dynamic spins and broad INS continuum[18-20]. However, according to theoretical predictions of the random-singlet state, the magnetic specific heat typically behaves as $C_m \sim T^\alpha$ with $0 < \alpha \leq 1$, and phonon thermal conductivity exhibits a nearly quadratic temperature dependence due to the resonant scattering by random singlets or frozen moments like in spin glasses[19], which is inconsistent with our observations of PrMgAl$_{11}$O$_{19}$.

In summary, we have measured specific heat and thermal conductivity of triangular-lattice Ising-type antiferromagnet PrMgAl$_{11}$O$_{19}$ single crystals. The magnetic specific heat down to 300 mK behaves as $C_m \sim T^2$ at 0 T, suggesting a $U(1)$ Dirac QSL ground state with gapless spinons. However, the thermal conductivity down to 80 mK reveals the absence of direct positive contribution from magnetic excitations, firmly excluding mobile gapless fermionic excitations. Instead, the observation of suppression temperatures $T_s$ at low fields indicates that phonons is strongly scattered by thermally-activated magnetic excitations out of a gap. The spin-phonon scattering rate can be well described as $1/\tau_{sp} \sim T\exp(-\Delta/T)$, implying the presence of gapped fermionic spinons. The spin gap is tiny at 0 T ($\Delta \sim 0.153$ K) and exhibits a linear increasing with applied field. Combining specific heat and thermal conductivity results, we identify the true ground state of PrMgAl$_{11}$O$_{19}$ as a gapped $Z_2$ QSL descended from $U(1)$ Dirac QSL, the instability of which may stem from the Ising anisotropy. Our study highlights PrMgAl$_{11}$O$_{19}$ as a promising QSL candidate and will promote the understanding of spin-1/2 TLIAF.

**Note added.** During writing this paper, we were aware of a parallel thermal conductivity study of

PrMgAl$_{11}$O$_{19}$ single crystals by Li *et al.*[50]. The observation of vanishing small residual linear term $\kappa_0/T$ at zero field and the field dependence of $\kappa$ are consistent with our work.

## Methods

**Sample preparation and characterization.**

High-quality single crystals of PrMgAl$_{11}$O$_{19}$ were grown by the floating zone method as described in ref. [48]. The large natural surface was identified to be the (00*l*) plane by using a D8 Advance X-ray diffractometer (XRD) from Bruker, as illustrated in Fig. 1c. The in-plane orientation was determined via Laue diffraction. The dc susceptibility measurements down to 1.8 K were performed in a magnetic property measurement system (MPMS, Quantum Design).

**Specific heat measurements.** The specific heat down to 0.3 K was measured on a piece of single crystal with 5.7 mg by the relaxation method in a physical property measurement system (PPMS, Quantum Design) equipped with a $^3$He cryostat.

**Thermal transport measurements.** The PrMgAl$_{11}$O$_{19}$ single crystal for the thermal conductivity measurements was cut and polished into a rectangular shape of dimensions 2.86 × 0.39 mm$^2$ in the *ab* plane, with a thickness of 0.32 mm along the *c* axis. The thermal conductivity was measured in a dilution refrigerator, using a standard four-wire steady-state method with two RuO$_2$ chip thermometers, calibrated *in situ* against a reference RuO$_2$ thermometer. The heat current was along the *a* axis, while the magnetic fields were applied along the *c* axis.

## Data availability

The data that support the findings of this study are available from the corresponding authors upon reasonable request.

## Acknowledgements

This work was supported by the National Key R&D Program of China (Grant No. 2022YFA1402203), the National Science Foundations of China (Grants No. 12034004 and 12204160), the Shanghai Municipal Science and Technology Major Project (Grant No. 2019SHZDZX01), the Natural Science Foundation of Shanghai (No. 23ZR1404500), and the Innovation Program for Quantum Science and Technology (Grant No. 2024ZD0300104).

## Author Contributions

S.Y.L. conceived the idea and designed the experiments. C.P.T. performed the specific heat and bulk thermal transport measurements with help from H.R.W., Y.H.J., D.Z.D.. Z.M. synthesized the single crystal samples. C.P.T. and S.Y.L. wrote the manuscript with comments from all authors. C.P.T. and Z.M. contributed equally to this work.

## Competing interests

The authors declare no competing interests.

## Additional Information

**Correspondence** and requests for materials should be addressed to S.Y.L. (shiyan_li@fudan.edu.cn).

Figure 1

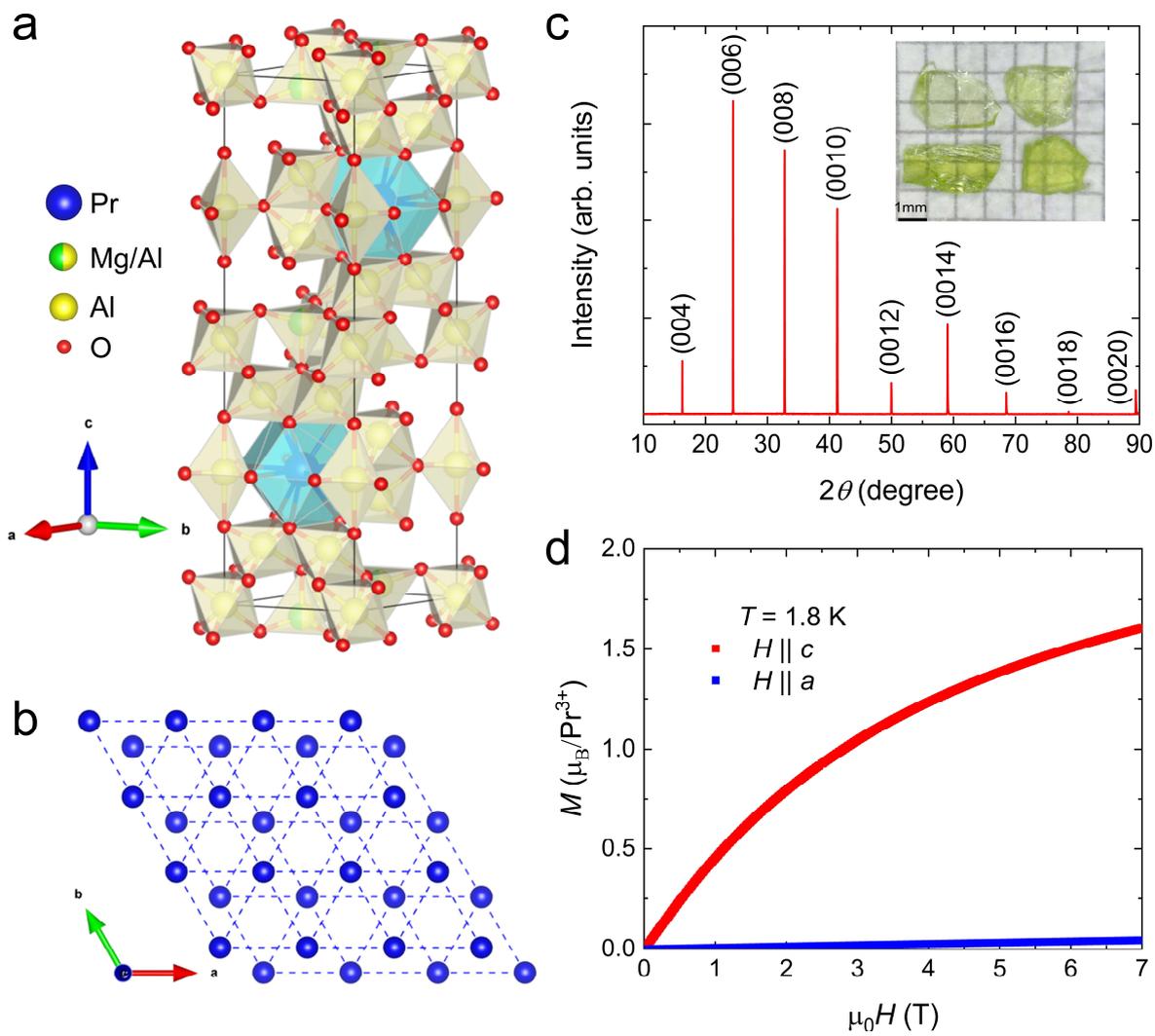

Figure 2

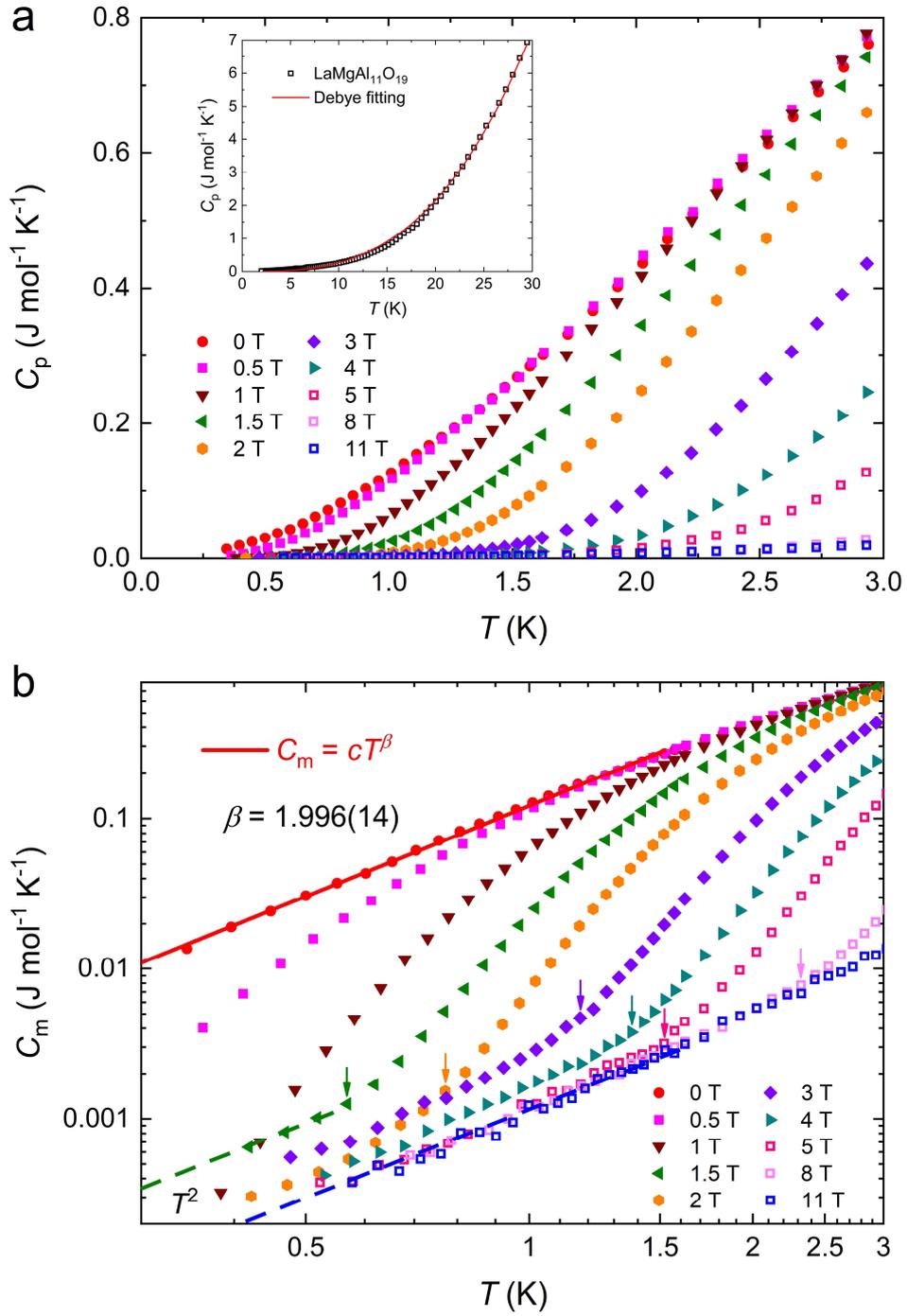

Figure 3

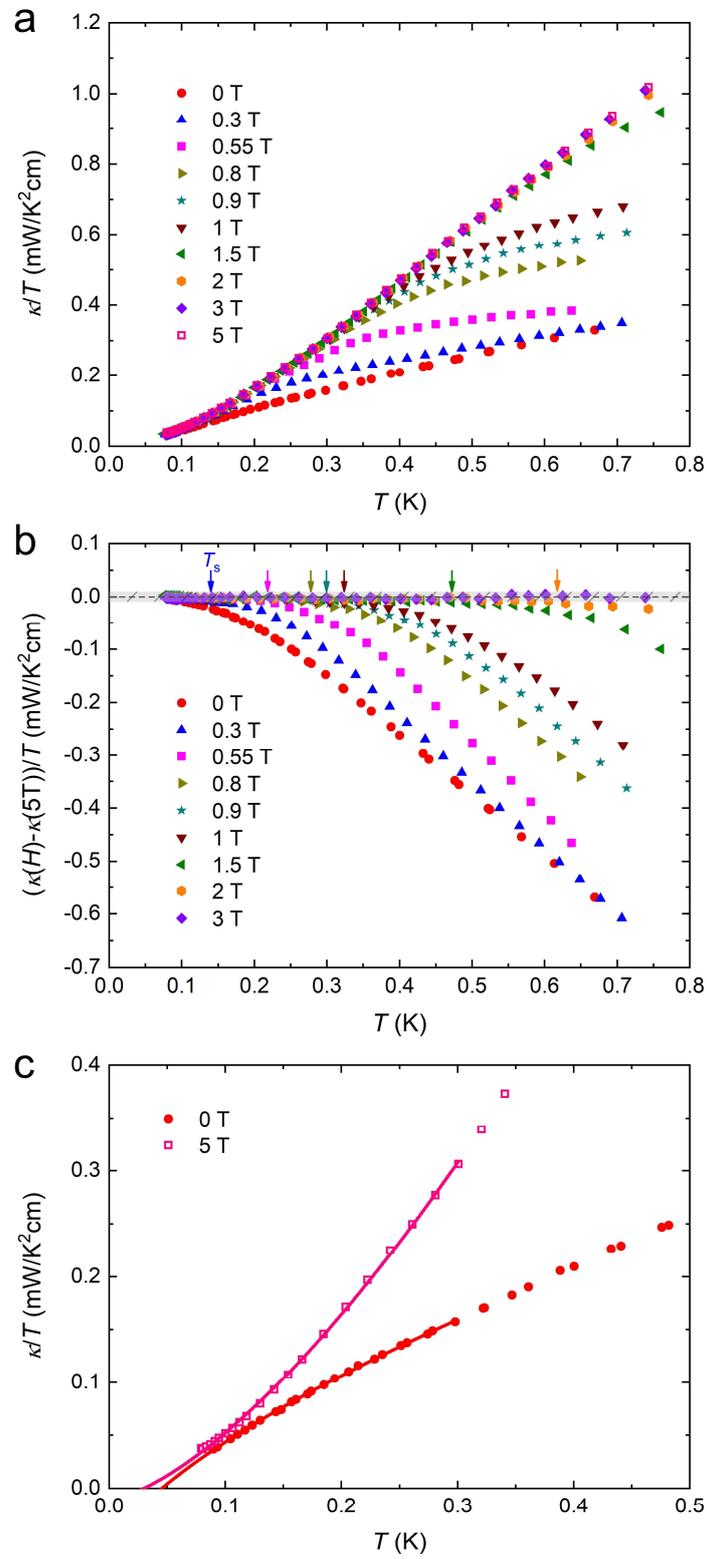

Figure 4

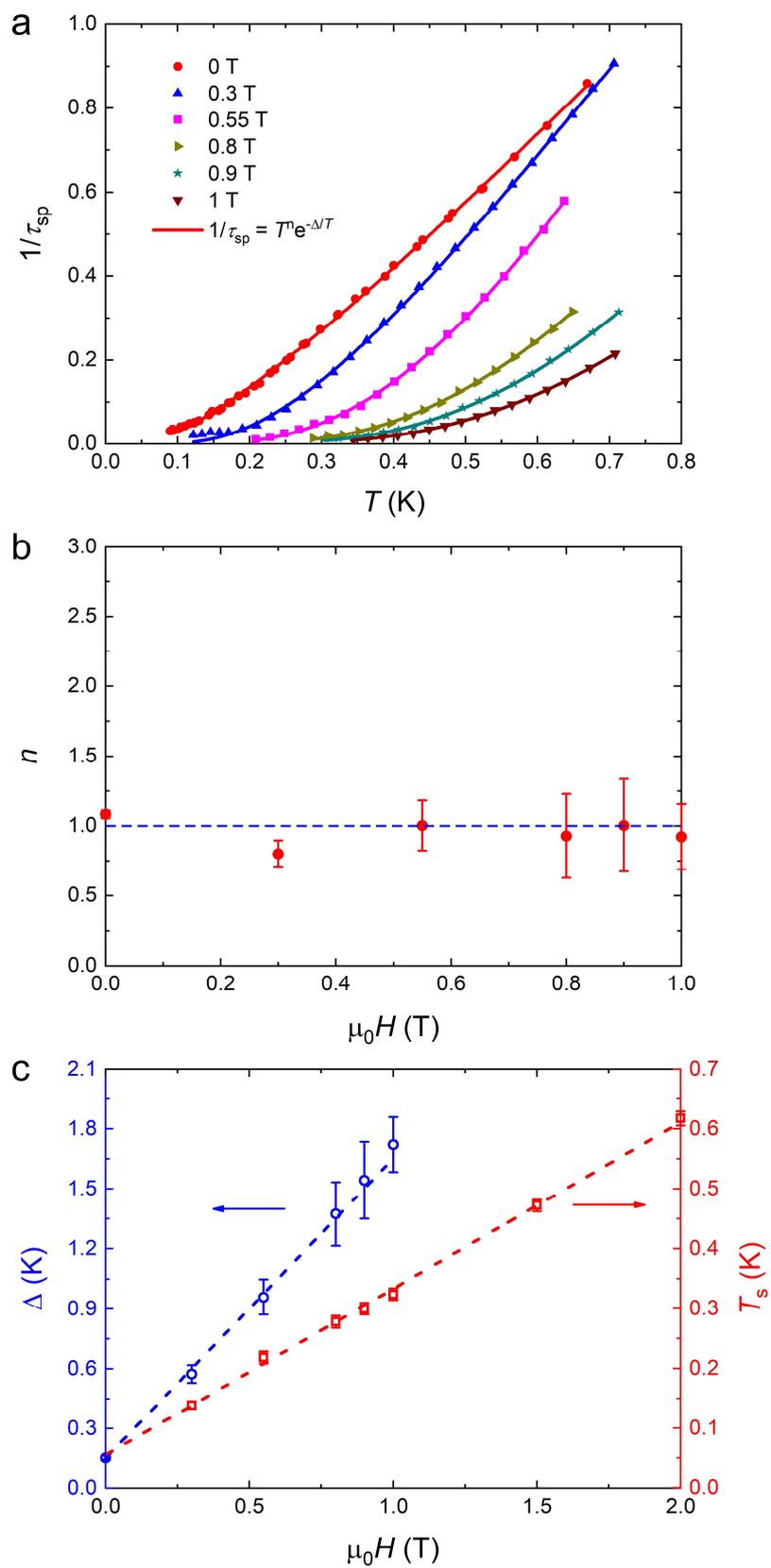

# Figure captions

**Figure 1 | Crystal structure and Ising anisotropy of PrMgAl$_{11}$O$_{19}$.** **a** Crystal structure of PrMgAl$_{11}$O$_{19}$. (Pr: blue; Mg: green; Al: yellow; O: red). PrO$_{12}$ tetrakaidecahedrons and Mg/AlO$_n$ polyhedrons (n represents the value of coordinate number) are presented by blue and yellow polyhedrons, respectively. **b** Top view of the Pr$^{3+}$ magnetic lattices within the *ab* plane. The Pr$^{3+}$ ions form triangular layers with an *ABAB* stacking arrangement. **c** Room-temperature X-ray diffraction pattern from the largest natural surface of a typical PrMgAl$_{11}$O$_{19}$ single crystal. Only (00*l*) Bragg peaks are found. Inset shows a photograph of PrMgAl$_{11}$O$_{19}$ single crystals. **d** Field dependence of dc magnetizations at 1.8 K with applied fields parallel to *c* and *a* axis, respectively. The great anisotropy reflects Ising-type exchange interaction in PrMgAl$_{11}$O$_{19}$.

**Figure 2 | Specific heat of PrMgAl$_{11}$O$_{19}$.** **a** Low-temperature specific heat results of PrMgAl$_{11}$O$_{19}$ single crystal in various magnetic fields. The inset shows the zero-field specific heat of LaMgAl$_{11}$O$_{19}$ polycrystalline sample between 2 and 30 K, which can be well fitted to Debye model $C_p \sim T^3$. **b** Log-log plot of magnetic specific heat of PrMgAl$_{11}$O$_{19}$, after subtracting the zero-field specific heat data of LaMgAl$_{11}$O$_{19}$ as phonon background. The solid red line shows a fitting to $C_m = cT^\beta$ below 1.5 K at 0 T, which yield $\beta = 1.996 \pm 0.014$, confirming a quadratic temperature dependence. Above 1.5 T, the low-temperature magnetic specific heat results return to a nearly $T^2$ dependence, represented by the dashed lines. The corresponding switching temperatures are denoted by arrows.

**Figure 3 | Thermal conductivity of PrMgAl$_{11}$O$_{19}$.** **a** Thermal conductivity data of PrMgAl$_{11}$O$_{19}$ single crystal in zero field and finite fields up to 5 T. **b** The negative contribution of magnetic excitations on the thermal conductivity is reflected in $\Delta(\kappa/T) \equiv (\kappa(H) - \kappa(5\ T))/T$. Overlapped regions are clearly shown below $T_s$ marked by arrows, above which bifurcations happen, indicating that the phonons are scattered by thermally-activated magnetic excitations out of a gap. The suppression temperature $T_s$ is defined when $\Delta(\kappa/T)$ starts to exceed $\pm 0.01$ mW K$^{-2}$ cm$^{-1}$, twice our experimental error bar, which is represented by the grey shaded region. **c** Thermal conductivity data for 0 T and 5 T. No apparent

overlapped region with $T_s$ is observed in the temperature range of our measurement. The solid lines represent fittings to $\kappa/T = a + bT^{\alpha-1}$ below 0.3 K.

**Figure 4 | Gapped quantum spin liquid ground state evidenced by spin-phonon scattering rate. a** Temperature dependence of spin-phonon scattering rate $1/\tau_{sp}$ at low fields. The solid lines represent fittings to $1/\tau_{sp} \sim T^n \exp(-\Delta/T)$ above their corresponding suppression temperatures $T_s$. **b** The obtained power $n$ with error bar, which is close to unity at all fields. **c** Field evolution of the obtained spin gap $\Delta$ and the suppression temperature $T_s$, both of which exhibit a linear field dependence represented by the dashed lines. At 0 T, the spin gap is estimated to be $\Delta \sim 0.153 \pm 0.010$ K, and the suppression temperature is extrapolated to be $T_s \sim 0.055 \pm 0.005$ K, indicating a gapped $Z_2$ QSL ground state of $PrMgAl_{11}O_{19}$. Note that the size of $\Delta$ is about 4 to 5 times of $k_B T_s$.